# Calculation of Forces on the High Granularity Calorimeter Stainless Steel Absorber Plates in the Compact Muon Solenoid Magnetic Field

Vyacheslav Klyukhin[1,2] on behalf of the CMS Collaboration

[1*]Skobeltsyn Institute of Nuclear Physics, Lomonosov Moscow State University, RU-119991, Moscow, Russia
[2]CERN, CH-1211, Geneva 23, Switzerland

[*]Corresponding author. E-mail : Vyacheslav.Klyukhin@cern.ch; ORCID: 0000-0002-8577-6531

**Abstract**

The general-purpose Compact Muon Solenoid (CMS) detector at the Large Hadron Collider (LHC) at CERN includes the hadronic calorimeter to register the energies of the charged and neutral hadrons produced in the proton-proton collisions at the LHC at a centre of mass energy 13.6 TeV. The calorimeter is located inside the superconducting solenoid of 6 m in diameter and 12.5 m in length creating the central magnetic flux density of 3.8 T. For operating optimally in the high pileup and high radiation environment of the High Luminosity LHC, the existing CMS endcap calorimeters will be replaced with a new high granularity calorimeter (HGCal) comprising of an electromagnetic and a hadronic section in each of the two endcaps. The hadronic section of the HGCAL will include 44 stainless steel absorber plates with a relative permeability value well below 1.05. The volume occupied by 22 plates in each endcap is about 21 m$^3$. The calculation of the axial electromagnetic forces on the absorber plates is a crucial element in designing the mechanical construction of the device. With a three-dimensional computer model of the CMS magnet, the axial forces on each absorber plate are calculated and the dependence of forces on the central magnetic flux density value is presented. The method of calculation and the obtained results are discussed.

## 1 Introduction

The Compact Muon Solenoid (CMS) multi-purpose detector [1] at the Large Hardon Collider (LHC) [2] registers the charged and neutral particles created in the proton-proton collisions at a center of mass energy 13.6 TeV. The detector includes a wide-aperture superconducting thin solenoid coil [3] with a diameter of 6 m, a length of 12.5 m, and a central magnetic flux density $B_0$ of 3.81 T created by the operational direct current of 18.164 kA. Inside the superconducting coil around the interaction point of proton beams the major particle subdetectors are located: a silicon pixel and strip tracking detectors to register the charged particles; a solid crystal electromagnetic calorimeter to register electrons, positrons and gamma rays; a barrel and endcap hadronic calorimeters of total absorption to register the energy of all the hadron particles. Outside the solenoid coil the muon spectrometer chambers register muon particles escaping the calorimeters.

In the next to a present run of the LHC, the High Luminosity (HL) operational phase is scheduled after a long shutdown of the machine. It is planned to level the instantaneous luminosity at $5 \times 10^{34}$ cm$^{-2}$ s$^{-1}$ with the goal of integrating some 3000 fb$^{-1}$ by the mid-2030s. The HL-LHC will integrate ten times more luminosity than the LHC, posing significant challenges for radiation tolerance and event pileup on detectors, especially for calorimetry in the forward region.

For optimal operation in the high pileup and high radiation environment of the HL-LHC the existing CMS plastic scintillator-based hadron endcap calorimeter will be replaced by a new High Granularity Calorimeter (HGCal) containing the silicon sensors and plastic scintillators as active material and stainless-steel absorber plates in the hadronic compartment [4]. The slightly magnetic stainless steel plates occupy at each endcap a volume about 21 m$^3$. Assuming to be produced with a relative permeability $\mu_{rel}$ well below 1.05 they nevertheless attract in the magnetic field of 3.81 T to the center of the CMS superconducting solenoid with substantial electromagnetic axial forces. With a three-dimensional (3D) CMS magnet computer model [5] based on a 3D finite-element code TOSCA (two scalar potential method) [6], developed in 1979 [7] at the Rutherford Appleton Laboratory, the axial force on each absorber plate is calculated for $\mu_{rel}$ equal to 1.05 and scaled to the conditions expected at the $B_0$ of 3.81 T using the measured dependence of $\mu_{rel}$ on the external magnetic flux density. Earlier, the influence of the HGCal stainless steel absorber plates with an extreme $\mu_{rel}$ value of 1.05 onto the CMS solenoid inner magnetic field in the location of the pixel and strip tracking detectors was investigated with similar modelling [8].

The article is organized as follows: Section 2 describes the model of the HGCal stainless steel absorber plates inside the CMS superconducting solenoid; Section 3 contains results of the axial magnetic force calculations as onto each stainless steel absorber plate as onto the entire endcap compartment; Section 4 presents a discussion of the obtained results and, finally, conclusions are drawn in Section 5.

## 2 Description of the HGCal hadronic compartment model

In Figure 1 a perspective view of the CMS detector is displayed with existing electromagnetic and hadronic endcap calorimeters colored with green and yellow colors, respectively. The calorimeters are shown at one side of the solenoid coil. At another side their positions are symmetrical with respect to the detector middle plane. For the HL-LHC operations both endcap calorimeters will be replaced with the HGCal endcaps.

To absorb the hadronic particles along the entire length of each HGCal endcap, 22 stainless steel disks with sicknesses from 45 to 95.4 mm interleaved with the air gaps of 21.55 mm form in each endcap the shapes shown in Figures 2 and 3. In the CMS magnet model, these disks are located at the distances from 3.6098 to 5.2239 m from the coil middle plane on both sides of the coil. These distances consider the shifts by 12 mm directed to the coil center on both sides according to the magnet yoke deformation under the magnetic forces when the magnet is switched on.

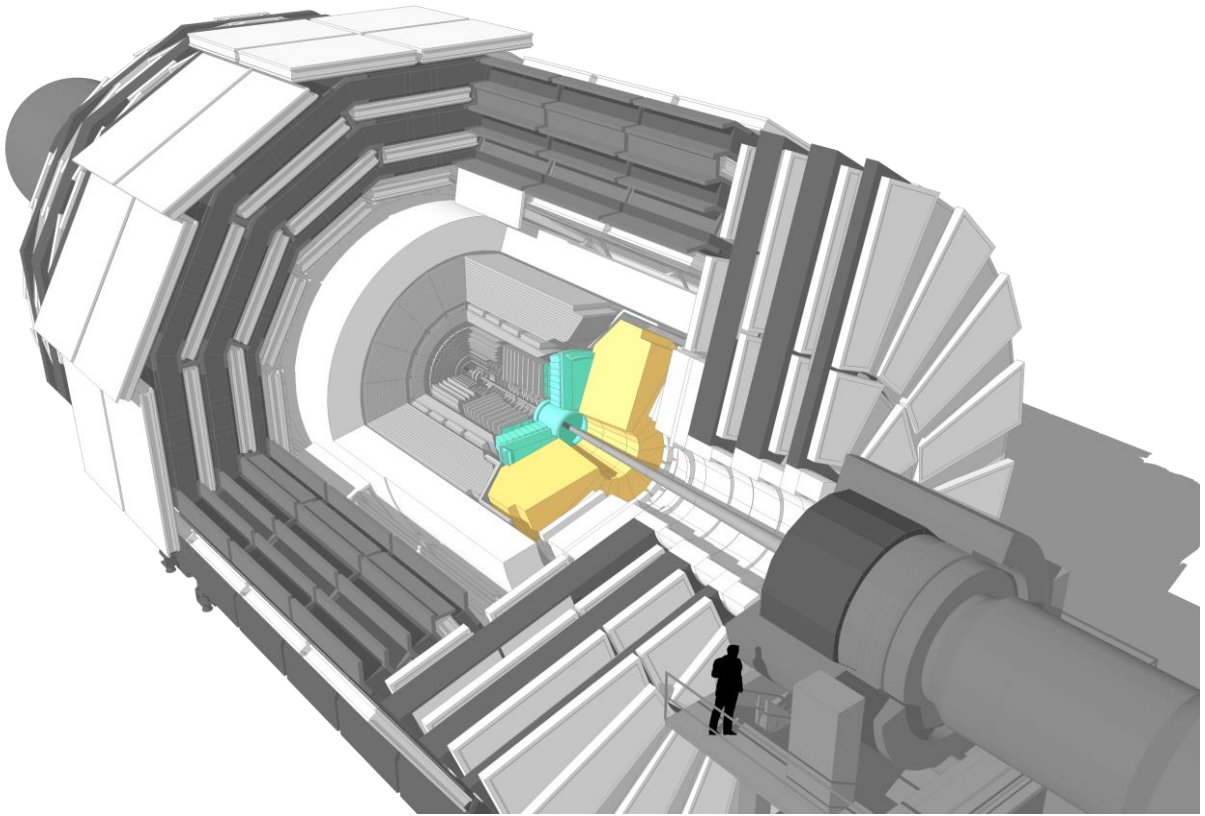

**Figure 1.** Perspective view of the CMS detector with 1/8 part opened. Inside this part an existing electromagnetic endcap calorimeter is shown with green color, and an existing hadronic endcap calorimeter is shown with yellow color. Positions of both endcap calorimeters are symmetrical with respect of the detector middle plane, and both calorimeters will be replaced with the HGCal.

Figures 2 and 3 are prepared with the updated latest CMS magnet 3D model [5]. The model is calculated with the operational direct current of 18.164 kA and with the stainless steel $\mu_{rel}$ value of 1.05 that corresponds to the limitation of the axial magnetic forces onto the absorbers. As shown in Figures 2 and 3, the absorber disks form two cones (small and large) and two cylinders at each side of the coil inner volume. The smallest outer diameter of the small cone is 3.2934 m, the outer diameter of the large cylinder is 5.2492 m, and the volume of the absorber plates modelled in each HGCal endcap is 18.9 m$^3$. The small cone has an angle of 19.09° with respect to the coil axis, and a similar angle of the large cone is 52.67°. An angle in transition disks ±5 between the small and large cones is 44.93°. In each endcap the first eight disks in the conical part (with numbers from ±1 to ±8) have an inner bore with a radius of 0.3136 m; the next four disks (with numbers from ±9 to ±12) have an inner bore with a radius of 0.3836 m; the last two disks in the conical part (with numbers ±13 and ±14) and all the eight cylinder disks (with numbers from ±15 to ±22) have an inner bore of 0.4448 m. A positive numeration of the disks runs from the smallest disk to a back disk in positive Z direction, a negative numeration of the disks runs in the same way in negative Z direction shown in Figures 2 and 3.

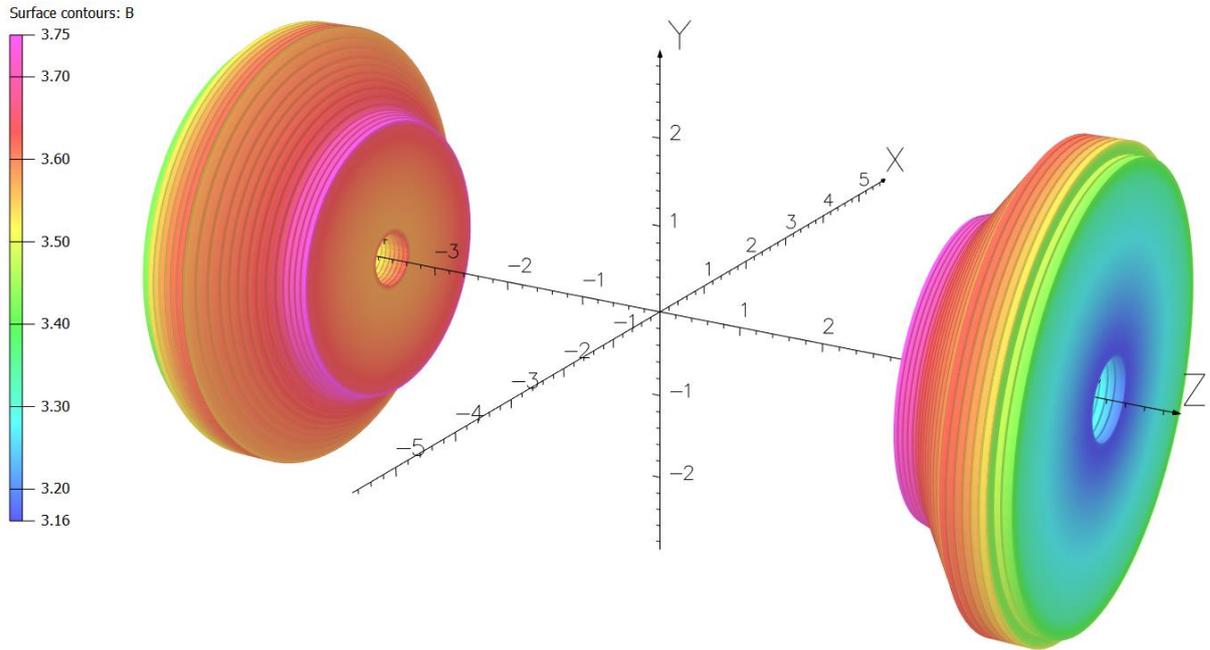

**Figure 2.** Perspective view of the stainless steel absorber disks in both HGCal endcaps. The color scale describes a distribution of the total magnetic flux density $B$ in Tesla on the disk surfaces. The maximum value of $B$ is 3.75 T, the minimum value is 3.16 T, the color scale unit is 0.1 T. The values on the coordinate axes are given in meters.

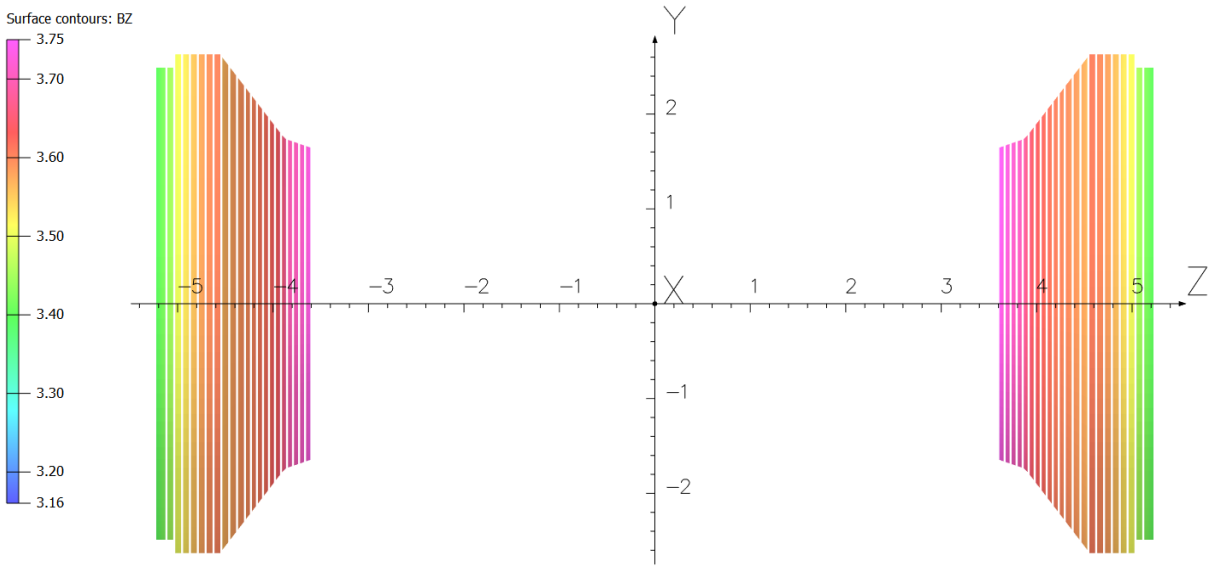

**Figure 3.** HGCal stainless steel absorber plates in the longitudinal section of the CMS coil inner volume. The color scale of the axial magnetic flux density $B_z$ on the disk surfaces is in Tesla with a unit of 0.1 T. The maximum value of $B_z$ is 3.75 T, the minimum value is 3.16 T. The values on the coordinate axes are given in meters. At positive Z coordinates the plates are numbering from 1 to 22 from origin of the coordinate system to positive Z values. At negative Z coordinates the plates are numbering from −1 to −22 from origin of the coordinate system to negative Z values. Each endcap compartment contains 22 stainless steel disk plates with thicknesses of 45 (disk numbers ±1), 41.5 (disk numbers from ±2 to ±11), 60.7 (disk numbers from ±12 to ±21) and 95.4 (disk numbers ±22) mm. The air gaps between the disks are of 21.55 mm.

The coordinate axes shown in Figures 2 and 3 represent the CMS coordinate system where the origin is in the center of the superconducting solenoid, the $X$ axis lies in the LHC plane and is directed to the center of the LHC machine, the $Y$ axis is directed upward and is perpendicular to the LHC plane, the $Z$ axis makes up the right triplet with the *X* and *Y* axes and is directed along the vector of magnetic flux density created on the axis of the superconducting coil.

In Figure 2, a distribution of the magnetic flux density total component *B* on the absorber plate surfaces is shown. A distribution of the axial magnetic flux density component $B_z$ is displayed in Figure 3. Both *B* and $B_z$ components have a maximum value of 3.75 T, and a minimum value of 3.16 T. The radial component of the magnetic flux density $B_r$ has an extremum value of ±0.27 T on the surfaces of the 95.4 mm thick back disks ±22.

The central magnetic flux density $B_0$ is equal to 3.817 T and is 0.21 % greater than that value in the present CMS configuration. Thus, at $\mu_{rel}$ value of 1.05, a contribution of the stainless steel plate magnetization into the CMS inner field is extremely small.

In Figure 4 the calculated axial magnetic flux density $B_z$ is shown along the *Z* coordinates at an *X*-coordinate equal to 1.259675 m. This *X*-coordinate value corresponds to averaging the halves of distances between the outer and inner disk radii of the first and the last disks in the endcap at positive *Z* coordinates. The $B_z$ values are calculated in three models of the CMS magnet: in Model 1 where the relative permeability inside the stainless steel absorber plates is equal to 1.05; in Model 2 where the relative permeability inside the absorber plates is equal to permeability of air, i. d. $\mu_{rel} = 1.0$; in Model 3 that corresponds to the present configuration of the CMS magnet without the absorber plates. Insertion of the absorber plates required the modification of the spatial finite-element mesh in Models 1 and 2 comparing with Model 3.

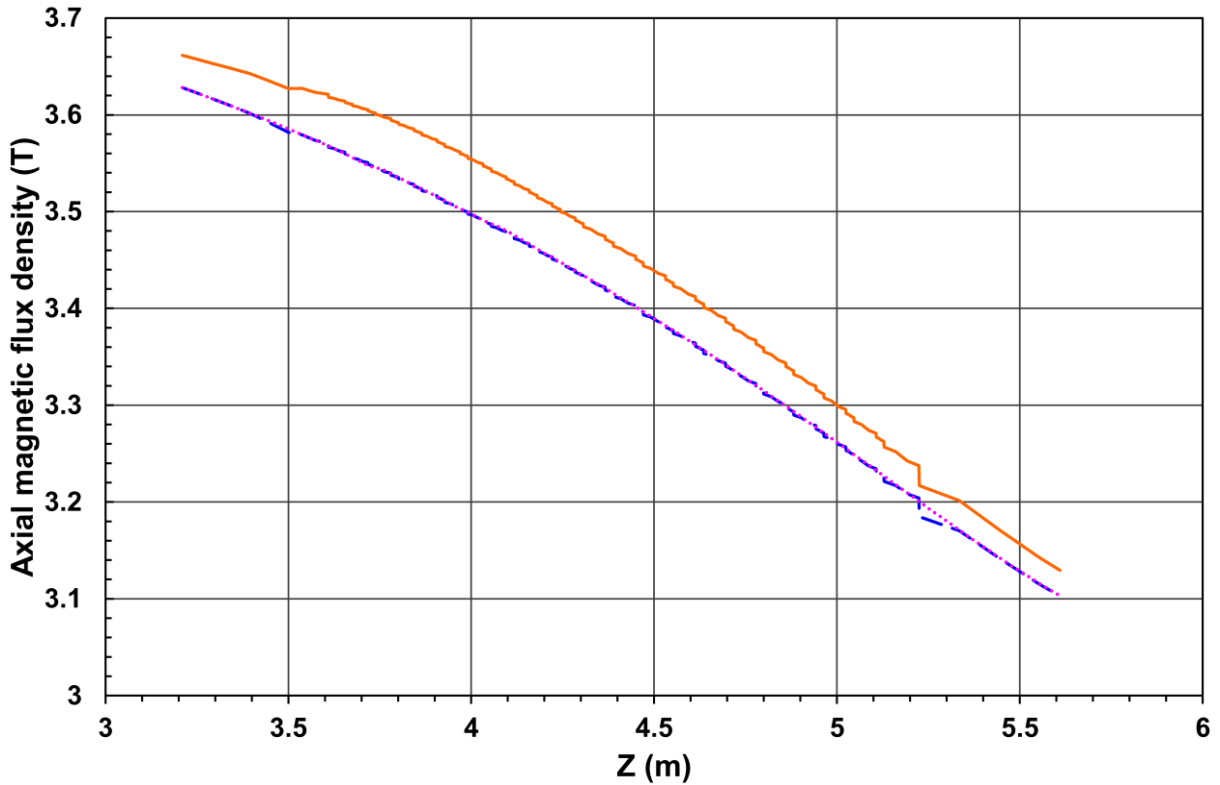

**Figure 4.** The axial magnetic flux density vs. positive *Z* at the *X*-coordinate equal to 1.259675 m in three different models of the CMS magnet. A smooth line corresponds to Model 1 with $\mu_{rel}$ value of 1.05 in the absorber plates, a dashed line corresponds to Model 2 with $\mu_{rel}$ value of 1.0 in the absorber plates, and a dotted line corresponds to Model 3 of the CMS magnet without the absorber plates.

As is visible in Figure 4, the $B_z$ values in Model 2 with $\mu_{rel} = 1.0$ in the absorber pates coincide with the $B_z$ values in Model 3 without the absorbers. This confirms the absence of the modified spatial mesh influence on the calculated results.

The curve of $B_z$ vs. *Z* in Model 1 with $\mu_{rel} = 1.05$ is rather smooth. It confirms a continuity of the axial magnetic flux density on the stainless steel – air interface since at *X* = 1.259675 m the axial magnetic flux density vector is normal to the surfaces of disks.

In Figure 5 the calculated axial magnetic field strength $H_z$ is shown along the *Z* coordinates at the *X*-coordinate equal to 1.259675 m. Here the $H_z$ values in Model 2 with $\mu_{rel} = 1.0$ in the absorber pates coincide with the $H_z$ values in Model 3 without the absorbers, and in Model 1 the axial magnetic field strength at the stainless steel – air interface is 1.05 times larger than inside the absorber plates that follows from the axial magnetic flux density continuity.

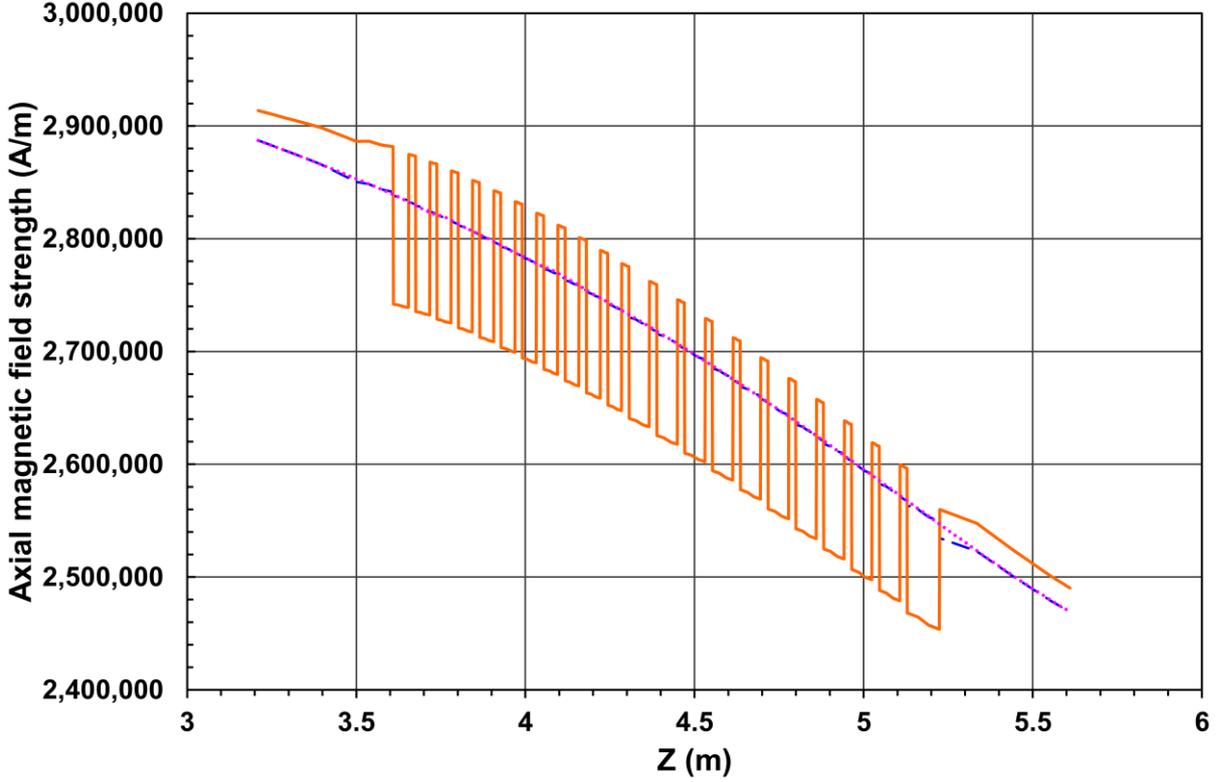

**Figure 5.** The axial magnetic field strength vs. positive Z at the X-coordinate equal to 1.259675 m in three different models of the CMS magnet. A smooth line corresponds to Model 1 with $\mu_{rel}$ value of 1.05 in the absorber plates, a dashed line corresponds to Model 2 with $\mu_{rel}$ value of 1.0 in the absorber plates, and a dotted line corresponds to Model 3 of the CMS magnet without the absorber plates.

## 2 Calculation of the axial electromagnetic forces

The force ***F*** acting to the stainless steel absorber plate with surface $S$ located in the magnetic field with the magnetic flux density ***B*** is calculated by integrating the Maxwell stress tensor [9] over the absorber surface. The following formula is valid for the integration in the air near the absorber surface:

$$\boldsymbol{F} = \frac{1}{\mu_0} \int_S \left[ \boldsymbol{B}\,(\boldsymbol{B} \cdot \boldsymbol{n}) - \frac{B^2}{2} \boldsymbol{n} \right] dS, \quad (1)$$

where $\mu_0$ is a permeability of free space and ***n*** is a unit vector of the outer normal to the integration surface $S$. It was observed that in the range of $\mu_{rel} < 1.1$ the calculated axial force $F_z$ is linear with $\mu_{rel}$ but has non-zero contribution at $\mu_{rel} = 1.0$ depending on the calculation conditions. If the region in the CMS magnet model, where the stainless steel absorbers are located, is described by a reduced scalar magnetic potential [5], then this contribution is negative. If the region with absorbers is described by a total scalar magnetic potential [5], then this contribution is positive. In both cases a slope of $F_z$ on $\mu_{rel}$ is approximately the same. It was also noticed that dividing the volume of the endcap compartment into a larger number of absorber plates somewhat increases the slope of linear dependence of $F_z$ on $\mu_{rel}$. Thus, in contrast to previous calculations [8] performed with five absorber sections in each endcap compartment located in the reduced scalar magnetic potential, in the present force calculations all 22 absorbers in each endcap compartment located in the total scalar magnetic potential are used.

The force calculation procedure includes calculating the axial forces in Model 1 using Eq. (1) and subtracting from the obtained values the systematic error calculated according to Model 2 with setting the relative permeability of air for the absorber plates. Figure 6, where the axial forces are calculated for each absorber plate in positive Z coordinates, illustrates this procedure. The form of the resulting curve in Figure 6 is explained in Section 4 containing the discussion of results.

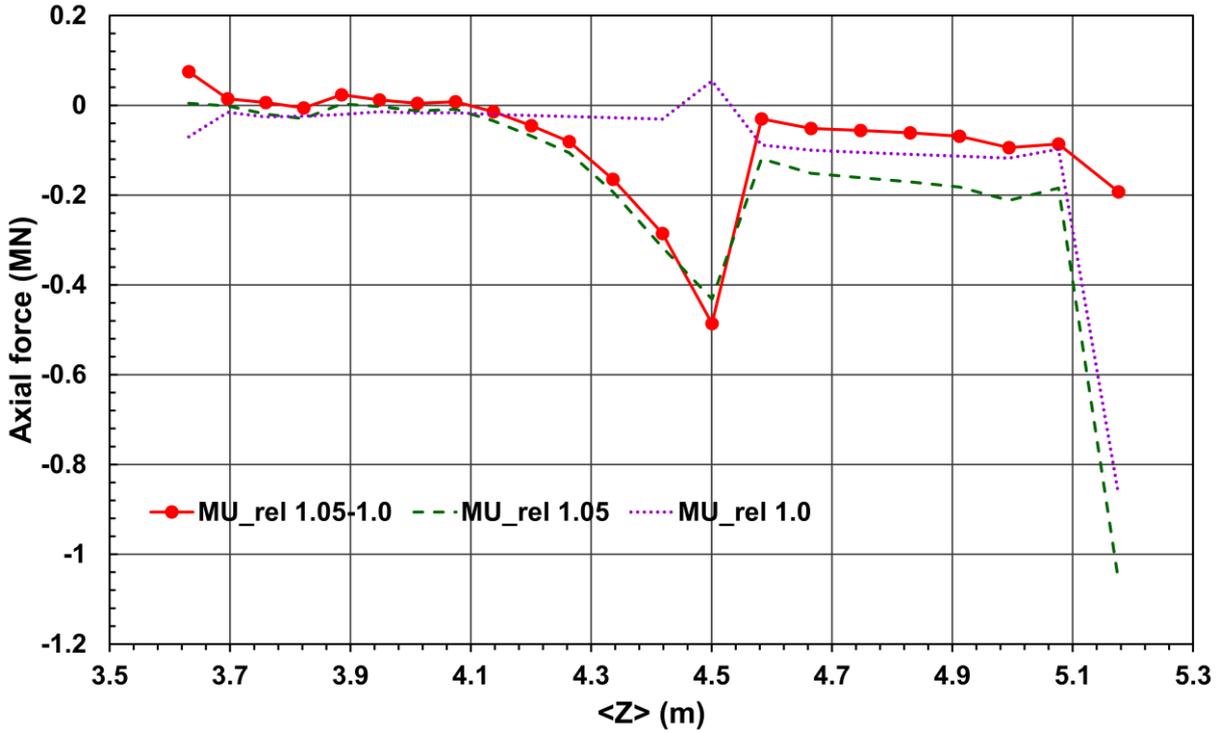

**Figure 6.** Axial forces (smooth line) for each HGCal absorber disk resulting from force values calculated with the relative permeability of 1.05 (dashed line; overestimated) and of 1.0 (dotted line; systematic error) at the central magnetic flux density of 3.81 T. Here <Z> means the Z coordinates of the median planes of the disks, marked with filled circles.

With $\mu_{rel}$ = 1.05 the resulting axial forces at the CMS coil operating current of 18.164 kA are −1.584 MN and +1.578 MN on the endcap compartment at positive and negative Z coordinates, respectively. These values calculated with the total scalar magnetic potential in the compartment regions are larger than those calculated earlier [8] with the reduced scalar magnetic potential in the same regions.

The value $\mu_{rel}$ = 1.05 is the limit value corresponding to the mechanical strength limit when an axial force acts on the endcap compartment integrated in the CMS magnet yoke. Under this limitation, the goal is to select a stainless steel material having the lowest possible relative permeability in a balance with a cost of producing the material. In addition, to simplify the choice of material, it is necessary to investigate the dependence of the relative permeability of the stainless steel material on the magnetic field strength or magnetic flux density in the air around the material. This dependence was measured for three selected samples of stainless steel material.

Figure 7 shows this measurement for one of samples: S1. The relative permeability of stainless steel in sample S1 rapidly drops from 1.04279 at an external magnetic flux density of 0.101 T to 1.01006 at an external magnetic flux density of 0.999 T. The approximation from the last measured value at 0.999 T to the expected value at the operating coil current of 18.164 kA, is made over the last three measured points using a power function as displayed in Figure 7. The resulting dependence of $\mu_{rel}$ on the external magnetic flux density in the region $B_0$ from 0.808 to 3.81 T has the following form:

$$\mu_{rel} = 1.01 \cdot B_0^{-007}. \qquad (2)$$

From this approximation the relative permeability of stainless steel in sample S1 is 1.000588 at 3.809442 T.

Using this value, together with two values measured at 0.101 and 0.999 T, the dependence of the axial forces for each endcap compartment on the CMS central magnetic flux density $B_0$ is calculated using Models 1 and 2. Figure 7 shows the results of calculations for the endcap compartment at positive, $F_{z\,abs1}$, and negative, $F_{z\,abs2}$, Z coordinates. As a result, at the CMS coil operating current of 18.164 kA the axial force $F_{z\,abs1}$ is −18.63 kN, and $F_{z\,abs2}$ is +18.56 kN.

Thus, the expected axial forces at the CMS central magnetic flux density of 3.81 T are 85 times less than the axial forces calculated with the extreme value of the relative permeability of stainless steel $\mu_{rel}$ = 1.05.

Figure 8 shows the dependence of the axial forces on the absorber number for positive and negative Z coordinates. The forces are calculated at two values of the relative permeability of the stainless steel disks: the maximum of 1.05 and the approximate value of 1.000588 at 3.81 T.

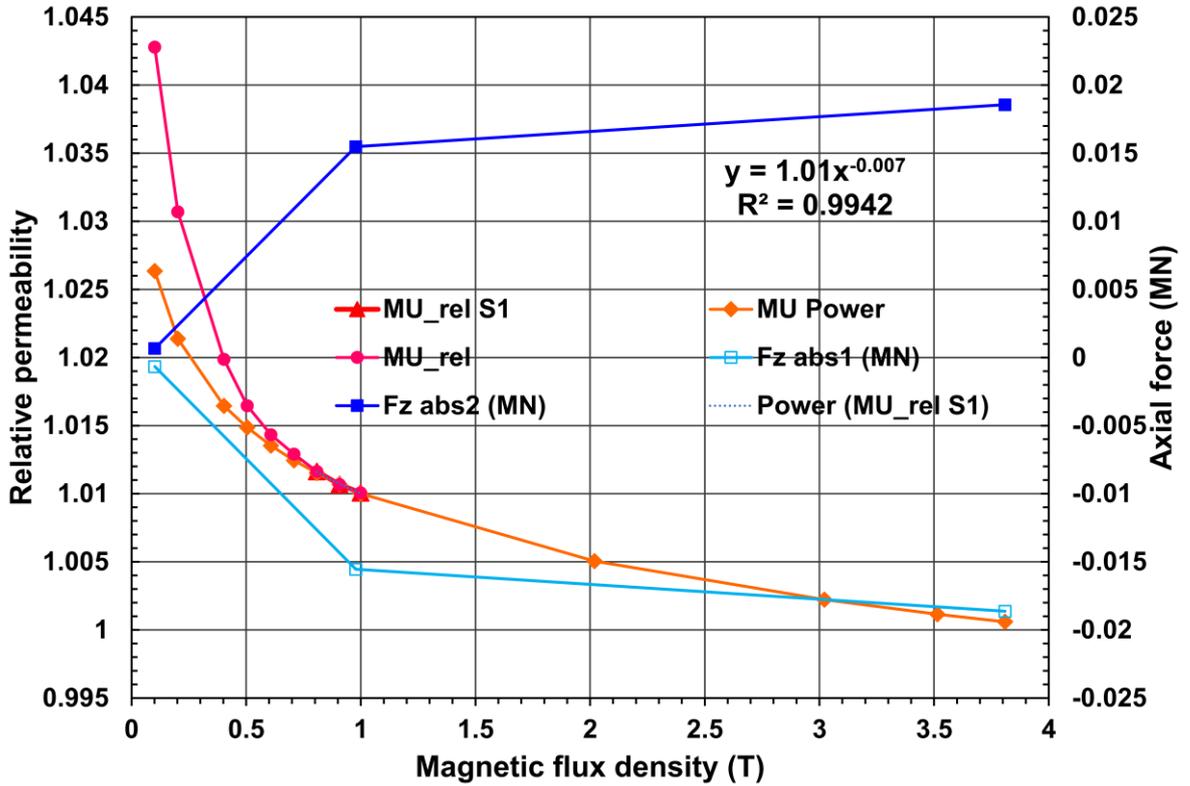

**Figure 7.** Left scale: Measured relative permeability (MU_rel S1 and MU_rel) in a stainless-steel sample S1 versus the applied magnetic flux density, approximated by a power function (MU Power). Right scale: Axial forces $F_z$ calculated for each endcap compartment at the different central magnetic flux density values inside the CMS coil. The relative permeability value approximated by a power function is 1.000588 at 3.81 T.

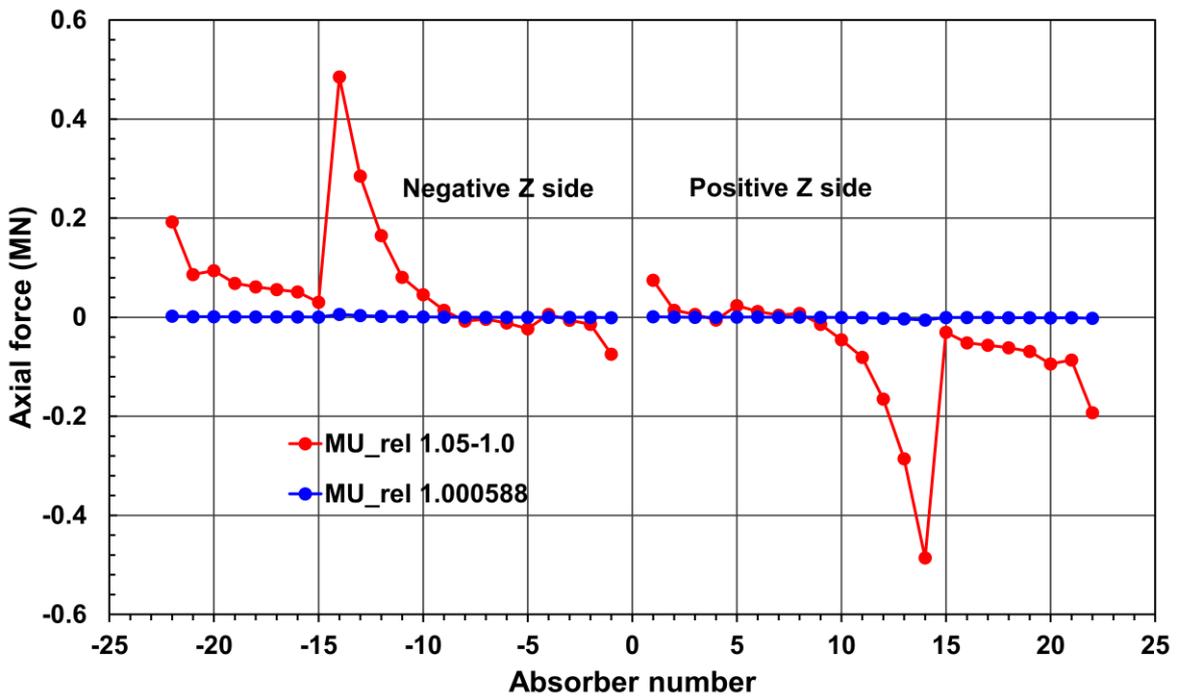

**Figure 8.** Axial forces on each HGCal stainless steel absorber disk calculated with two values of the relative permeability: the maximum of 1.05, and the approximate value of 1.000588 at 3.81 T. The forces calculated with $\mu_{rel}$ of 1.000588 are well below the safety margins.

## 4 Discussion

To explain the form of the dependence of the axial force on the $Z$ coordinate shown in Figure 6, or on the disk number as shown in Figure 8, consider the forces acting to the conical disk. In Figure 9, the conical disk at the positive $Z$ coordinates is shown in a longitudinal view. At the edge of the disk the surface of integration with Eq. (1) is divided by a dashed line into three parts: a cylinder part with the front plane at $Z_1$ and the back plain at $Z_2$, a conical part on the front side between $Z_1$ and $Z_2$; and an annular part at the back side at $Z_2$.

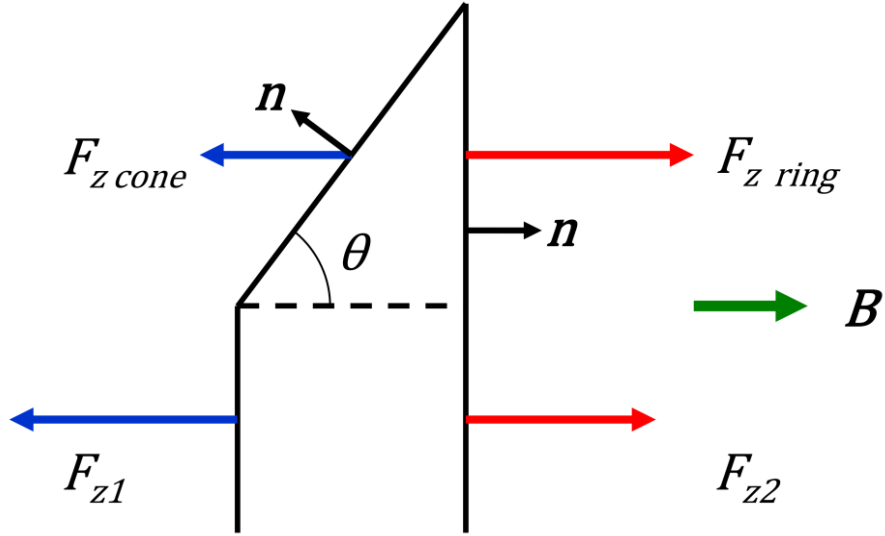

**Figure 9.** Scheme of the action of axial forces on a conical disk at the positive $Z$ coordinates. Here $B$ is a direction of the magnetic flux density; $n$ is a unit vector of the outer normal to the integration surface; $\theta$ is an angle of the cone relative to the CMS coil axis; $F_{z\,cone}$ is an axial force acting on the conical surface of the disk; $F_{z\,ring}$ is an axial force acting on the outer annular surface of the disk bounded by the dotted line; $F_{z1}$ and $F_{z2}$ are axial forces acting on the two central surfaces of the disk.

For these three surfaces, integration by Eq. (1) can be represented in three forms. The following equation expresses the axial force $F_{z\,cyl}$ acting on the central surfaces of the disk:

$$F_{z\,cyl} = -\frac{1}{\mu_0}\left\{\int_S \left(B_{z1}^2 - \frac{B_1^2}{2}\right)dS - \int_S \left(B_{z2}^2 - \frac{B_2^2}{2}\right)dS\right\}, \tag{3}$$

where index 1 corresponds to the front surface of the disk, index 2 corresponds to the back side of the disk and $S$ is the front or rear surface of the central cylinder. The axial force $F_{z\,cone}$ acting on the conical surface of the disk is described by the following equation:

$$F_{z\,cone} = -\frac{1}{\mu_0}\int_S \left[\frac{(2B_z^2 - B^2)\sin\theta}{2} - B_z B_r \cos\theta\right]dS, \tag{4}$$

where $B_r$ is the radial component of the external magnetic flux density and $\theta$ is an angle of the cone with respect to the $Z$ axis. The axial force $F_{z\,ring}$ acting on the outer annular surface of the disk is described by the following equation:

$$F_{z\,ring} = +\frac{1}{\mu_0}\int_S \left[B_z^2 - \frac{B^2}{2}\right]dS. \tag{5}$$

Eq. (3) describes the main contribution to the axial force. In the region of the positive $Z$ coordinates $F_{z\,cyl}$ is negative, since $B_z \sim B$ and $B_{z1} > B_{z2}$ which follows from Figure 4. In the region of negative $Z$ coordinates $F_{z\,cyl}$ is positive since the directions of the unit vectors of the outer normal to the front and back surfaces have opposite directions relative to those in the region of the positive $Z$ coordinates. In both cases the axial forces act towards the CMS coil center. Combining Eqs. (4) and (5) we conclude that the axial force on the peripheral part of the conical disk $-|F_{z\,cone}| + |F_{z\,ring}|$ is positive in the region of positive $Z$ coordinates and negative in the region of the negative *Z* coordinates, and in both cases the forces act in the direction from the center of the CMS coil. From these two circumstances, we can conclude that the shape of the axial force dependence on the *Z* coordinate or disk number shown in Figures 6 and 8, respectively, could be explained by a quadratic rise of the surfaces of the cylinder part of the conical disks and by a linear increase in the differences between the axial and total components of the magnetic flux densities on the front and back sides of the disk.

## 5 Conclusions

In this study, the calculation of the electromagnetic forces acting on the stainless steel hadronic compartments of a new high granularity calorimeter is performed using the Compact Muon Solenoid (CMS) magnet modified three-dimensional model. It is planned to place 22 stainless steel disks 45, 41.5, 60.7 and 95.4 mm thick of each compartment with 21.55 mm air gaps between the disks in the CMS magnetic field with a magnetic flux density of 3.81 T. With a relative permeability value of 1.05 the maximum axial force acting on the disk plate is 0.486 MN, and the maximum total axial force acting on 22 plates in the compartment is 1.584 MN. With a relative permeability of 1.000588, approximated for a central magnetic flux density of 3.81 T, the maximum axial force acting on the disk plate is 5.72 kN, and the maximum total axial force acting on the 22 plates in the compartment is 18.6 kN, which is 85 times less and well below the safety margins. The form of the dependence of the axial force on the position of the disk in the endcap compartment is determined by the increase in the surfaces of the disk with distance from the center of the CMS coil and by the distribution of the magnetic flux density along the radii of the disks in a non-uniform magnetic flux at the edges of the CMS coil.

## Declarations

**Supplementary Materials**: Not applicable.

**Author Contributions**: The author contributed to the study conception and design. Material preparation, data collection and analysis were performed by Vyacheslav Klyukhin. The first draft of the manuscript was written by Vyacheslav Klyukhin who read and approved the final manuscript.

**Funding:** This research received no external funding.

**Data Availability Statement:** No data is available.

**Acknowledgments:** The author is very grateful to Hubert Gerwig of CERN for initiating this research, as well as for several years of interest in the progressive results and fruitful discussions.

**Conflicts of Interest:** The authors declare no conflict of interest.

## References

[1] CMS Collaboration. The CMS experiment at the CERN LHC. *J. Instrum.* **2008**, *3*, S08004. https://doi.org/10.1088/1748-0221/3/08/S08004.
[2] Evans, L.; Bryant, P. LHC Machine. *J. Instrum.* **2008**, *3*, S08001. https://doi.org/10.1088/1748-0221/3/08/s08001.
[3] Hervé, A. Constructing a 4-Tesla large thin solenoid at the limit of what can be safely operated. *Mod. Phys. Lett. A* **2010**, *25*, 1647–1666. https://doi.org/10.1142/S0217732310033694.
[4] CMS Collaboration. The Phase-2 Upgrade of the CMS endcap calorimeter. Technical Design Report. *CERN-LHCC-2017-023, CMS-TDR-019*; CERN: Geneva, Switzerland. **2018**, pp. 11–20. ISBN: 978-92-9083-459-5. Available online: https://cds.cern.ch/record/2293646 (accessed on 1 October 2023).
[5] Klyukhin, V. Design and Description of the CMS Magnetic System Model. Symmetry **2021**, *13*, 1052. https://doi.org/10.3390/sym13061052.
[6] *TOSCA/OPERA-3d 18R2 Reference Manual*; Cobham CTS Ltd.: Kidlington, UK, 2018; pp. 1–916.
[7] Simkin, J.; Trowbridge, C. Three-dimensional nonlinear electromagnetic field computations, using scalar potentials. In *IEE Proceedings B Electric Power Applications*; Institution of Engineering and Technology (IET): London, UK, **1980**; *127*, 368–374. https://doi.org/10.1049/ip-b.1980.0052.
[8] Klyukhin, V.; on behalf of the CMS Collaboration. Influence of the high granularity calorimeter stainless steel absorbers onto the Compact Muon Solenoid inner magnetic field. *SN Appl. Sci.* **2022**, *4*, 235. https://doi.org/10.1007/s42452-022-05122-9.
[9] Tamm, I.E. Fundamentals of the theory of electricity; Ninth Russian Edition, Moscow: Nauka, 1976; pp. 385–387 (in Russian).